\documentclass[prd, preprint, 12pt]{revtex4-1}

\usepackage{amsmath}
\usepackage{amssymb}
\usepackage{setspace}
\usepackage{graphicx}
\usepackage{natbib}
\usepackage{float}

\begin{document}
 
 % ! TEX spellcheck
 %
%\ \vskip 1.0 in

\begin{center}
 { \large {\bf Spacetime Fluctuations and a \\  Stochastic Schr\"odinger-Newton Equation }}

%\smallskip

\vskip 0.3 in

{\large{\bf Sayantani Bera$^{*}$, Priyanka Giri$^{*,\dagger}$ and Tejinder P.  Singh$^{*}$}}

%{\it $^{*}$Indian Institute of Technology Bombay, Powai, Mumbai 400076, India}\\  
{\it $^{*}$Tata Institute of Fundamental Research,}
{\it Homi Bhabha Road, Mumbai 400005, India}\\
{\it $^{\dagger}$Department of Physics, National Institute of Technology, Agartala 799046, India}\\
\bigskip
{\tt sayantani.bera@tifr.res.in, priyanka.physics@nita.ac.in, tpsingh@tifr.res.in}\\

\end{center}

\centerline{\bf ABSTRACT}
\noindent We propose a stochastic modification of the  Schr\"{o}dinger-Newton equation which takes into account the effect of extrinsic spacetime fluctuations. We use this equation to demonstrate gravitationally induced decoherence of two gaussian wave-packets, and obtain a decoherence criterion similar to those obtained in the earlier literature in the context of effects of gravity on the Schr\"{o}dinger equation.

\bigskip
\noindent 
\noindent 

\vskip 1 in

\setstretch{1.4}

\section{Introduction}
\noindent Is the apparent collapse of the wave-function during a quantum measurement caused by a dynamical physical process which results from possible modification of the Schr\"{o}dinger equation? Or can it be explained within the framework of standard quantum theory via environmental decoherence and the many-worlds interpretation, or through a reformulation such as Bohmian mechanics? In the coming years it might become possible to decisively answer this question experimentally, thanks to advances in technology, and new innovative ideas for experiments based on optomechanics and interferometry \cite{RMP:2012}.

The focus of such experiments and ideas for experiments is to test dynamical collapse theories such as Continuous Spontaneous Localisation [CSL] which involve a stochastic nonlinear modification of the Schr\"{o}dinger equation.  CSL is a phenomenological theory with two free parameters, designed to solve the measurement problem, explain the Born probability rule, and to explain the apparent absence of superpositions of macroscopic states \cite{Ghirardi:86,Ghirardi2:90}. However, at the present state of understanding it is unclear as to what is the fundamental origin of CSL: why should there be a stochastic modification of the Schr\"{o}dinger equation? Possible explanations include the existence of a fundamental stochastic field in nature, which couples nonlinearly to matter fields and results in an anti-Hermitean modification to the Hamiltonian. Alternatively, quantum theory maybe a coarse-grained  approximation to a deeper theory such as Trace Dynamics, and stochastic modifications arise when one goes beyond the leading order  approximation. A third possible explanation is that gravity plays a role in bringing about collapse of the wave-function  \cite{RMP:2012, gao, Singh:2015}. 
 The present paper is concerned with a specific, modest aspect concerning  the possible role of gravity.

The idea that gravity plays a role in collapse of the wave-function has been around for the last fifty years, and has been pursued by many investigators starting with the works \cite{Karolyhazi:66,Karolyhazi:86, Karolyhazy:74, Karolyhazy:90, Karolyhazy:95, Karolyhazy:1982, Frenkel:77, Frenkel:90, Frenkel:95, Frenkel:2002, Frenkel:97},  and also pursued by Di\'osi and collaborators  \cite{Diosi:87, Diosi:07, Diosi:89, Diosi:87a}.  The basic principle behind the idea is easy to state and understand. Gravitational fields are produced by material bodies; and largely by macroscopic material bodies. However even macroscopic bodies are not exactly classical, and their position and momenta are subject to the uncertainty principle. It is plausible then [unless one invokes semiclassical gravity] that the gravitational field produced by these bodies is also subject to intrinsic fluctuations, which induce stochasticity in the space-time geometry, which cannot be ignored. Thus when one is studying the 
Schr\"{o}dinger evolution of a quantum system on a background spacetime (even a flat Minkowski spacetime), one can in principle not ignore these spacetime fluctuations. When one makes models to see how these fluctuations affect the standard Schr\"{o}dinger evolution, it is found [as should be the case] that microscopic objects are not affected by the gravitational fluctuations, so that the conventional picture of quantum theory and the linear superposition principle continues to hold for them. However, the 
Schr\"{o}dinger evolution of a macroscopic object is significantly affected, leading to gravitationally induced decoherence, thus providing at least a partial resolution of the measurement problem. While it has not been shown that collapse of the wave-function can be achieved through gravity, models strongly suggest that fundamental decoherence [loss of interference without loss of superposition] can be achieved through gravity [without the need for an environment]. It is hoped that when properly understood, gravity might be able to provide an underlying explanation for CSL.

One of the earliest pioneering works investigating gravity induced decoherence is due to Karolyhazy, who proposed that the quantum nature of objects imposes a minimum uncertainty [different from Planck length] on the accuracy with which length and time intervals can be measured. This is interpreted as an intrinsic property of spacetime, which is modelled as resulting from a stochastic metric perturbation having a [non-white] gaussian two-point correlation. The Schr\"{o}dinger evolution of a quantum object is modified to include the effect of this stochastic potential, and it is shown that gravitational decoherence can be achieved for macroscopic objects. This model has been studied further by Karolyhazy and collaborators. In a different model, Di\'osi  has modelled  the intrinsic quantum uncertainty of the Newtonian gravitational potential [resulting from the quantum nature of the probe] by a [white-noise] gaussian correlation, and again demonstrated gravitational decoherence. This model has also been studied further by various authors. The Karolyhazy model and the Di\'osi model have been recently compared in \cite{Bera:2015}. 

Models such as those of Karolyhazy and Di\'osi study the effect of extrinsic space-time fluctuations on the Schr\"{o}dinger equation. A different gravitational effect is due to the self-gravity of the quantum object: how does the Schr\"{o}dinger equation get modified by the gravity of the very particle for which this equation is being written? One possible way to describe this effect is to propose that to leading order the particle produces a classical potential satisfying the Poisson equation, whose source is a density proportional to the quantum probability density. The Schr\"{o}dinger equation is then modified to include this potential [a kind of back-reaction] and the modified equation is known as the Schr\"{o}dinger-Newton [SN] equation \cite{Penrose:96, Penrose:98, Penrose:00, Diosi:84}. The SN equation has been studied extensively in many papers, for its properties and possible limitations \cite{Bernstein:98,giulini2011gravitationally,Harrison:2003,Moroz:98,Ruffini:69,Giulini2012,Giulini2013,Hu2014,Anastapoulos:2014, Bahrami:2014, Colin2014, Derakhshani:2014}.
 One important feature of the SN equation is a gravitationally induced inhibition of dispersion of a wave-packet \cite{giulini2011gravitationally}.

However, the SN equation is not intended to explain gravitationally induced decoherence or collapse of the wave-function. It cannot achieve that because it lacks a stochastic feature, unlike the Karolyhazy and Di\'osi models, which employ a stochastic gravitational field in the Schr\"{o}dinger equation. The SN equation only incorporates the deterministic back-reaction of self-gravity in a semiclassical fashion, and one worrisome outcome of this deterministic nonlinearity is superluminal signalling. It is desirable to modify the SN equation into a stochastic equation, possibly by including higher order corrections to self-gravity, or otherwise. This brings home the possibility that the SN equation can take into account self-gravity as well as perhaps produce gravitational decoherence, though it remains to be seen whether the superluminal feature can be gotten rid of by including stochasticity.  

Another interesting aspect which seems worth considering, and which is the subject of the present paper, is to simultaneously take into account the effect of self-gravity and of extrinsic spacetime fluctuations. After all, that seems to be a rather natural and wholesome way of accounting for the role of gravity in Schr\"{o}dinger evolution. In this spirit, we write down, in the next section, a modified SN equation which includes a stochastic potential representing extrinsic spacetime uncertainty and having Di\'osi's white noise correlation. In Section III, we use this stochastic equation to demonstrate gravitational decoherence of two gaussian states of a free particle, and obtain decoherence criteria similar to those obtained by Diosi. In Section IV we discuss the implications of our results, and compare them with earlier work. Details of some of the integrals that appear in Section III, are given in Appendix I. 

\section{A new proposal for a stochastic Schr\"odinger-Newton Equation}
The SN equation for describing self-gravity effects is obtained by substituting the solution of potential $V$ from the semiclassical Poisson equation
\begin{equation}
\nabla^{2}V = 4\pi G m |\Psi|^2
\end{equation}
in the Schr\"{o}dinger equation
\begin{equation}
i\hbar\frac{\partial\Psi}{\partial t} = - \frac{\hbar^2}{2m} \nabla^{2}\Psi + mV \Psi
\end{equation}
so as to arrive at \cite{Penrose:96, Penrose:98, Penrose:00, Diosi:84}
\begin{equation}
i\hbar \frac{\partial \Psi({\bf r}, t)}{\partial t} =  - \frac{\hbar^2}{2m} \nabla^{2}\Psi ({\bf r}, t) -
Gm^2 \int \frac{ |\Psi ({\bf r'},t)|^2}{|{\bf r}-{\bf r'}|} d^3 r' \Psi ({\bf r},t)
\end{equation}

In contrast,  Di\'osi's model considers the effect of extrinsic spacetime fluctuations on the wave-function for the center-of-mass ${\bf x}$ of a matter distribution $f({\bf r}|X)$. This effect is modelled by including a stochastic potential $\phi({\bf r},t)$ in the Schr\"{o}dinger equation, so as to get:
\begin{equation}
i\hbar\dot{\Psi}({ X},t) = \left( \hat{H}_{0} + \int \phi ({\bf r'},t){f}({\bf r'}|X)\ d^3 r' \right) \Psi(X,t)
\label{stod}
\end{equation}
where ${f}({\bf r'}|X)$ stands for the local mass density of the system for a configuration
 $X$. 

The stochastic potential is described by gaussian white noise which has the two-point correlation
\begin{equation}
\langle\phi({\bf r}, t) \phi({\bf r}', t')\rangle = \hbar G |{\bf r} - {\bf r}'|^{-1} \ \delta(t-t')
\label{dcorr}
\end{equation}
We propose to construct a stochastic SN equation for considering the joint effect of self-gravity and the extrinsic spacetime fluctuations. This is thus a hybrid of the SN equation and the Di\'osi model, with one difference: the matter distribution ${f}({\bf r'}|X)$ in Di\'osi's model is now replaced by the quantum mass density 
$m|\Psi|^2$. 

In so doing, there however arise two subtleties, which can be pointed out more easily by first looking at Eqn. (\ref{stod}). Here, if the mass density  ${f}({\bf r}|X)$ is to be replaced by a quantum mass density, and the (linear) structure of the equation is to be preserved, then the wave function entering this mass density  must be the unperturbed wave function $\psi$ which is a solution of the free part of (\ref{stod}) which does not include the stochastic part. Only then can one reproduce the decoherence results of \cite{Diosi:87} as is borne out from the analysis below. The second subtlety is that when one replaces the distribution ${f}({\bf r}|X)$ by the unperturbed quantum probability density for a point particle, one must find an analog for the configuration variable $X$. For a classical matter distribution $X$ is the set of particle coordinates, and for a single point particle at coordinate ${\bf r}$, one has $X={\bf r}$ and $f({\bf r'|X)}=m\delta ({\bf r'}-{\bf r})$. The appropriate replacement for $X$ is that the unperturbed wave function 
$\psi$ depends parametrically on the position  variable ${\bf r}$ on which the full stochastic wave function $\Psi({\bf r})$ depends. Thus for the case of a quantum probability density replacing the distribution 
${f}({\bf r}|X)$, Eqn. (\ref{stod}) can be revised as
 \begin{equation}
i\hbar\dot{\Psi}({\bf r},t) = \left( \hat{H}_{0} + m\int |\psi_{\bf r}({\bf r'},t)|^2\; \phi ({\bf r'},t)\ d^3 r' \right) \Psi({\bf r},t)
\label{stod2}
\end{equation}
By $\psi_{{\bf r}}({\bf r'},t)$ we will specifically mean a gaussian wave-packet  which is a solution of the free equation (i.e. without the stochastic part) and which is peaked at ${\bf r}$. The choice of such a state exactly reproduces the decoherence results of \cite{Diosi:87} as we will see below.

In view of the above, we propose to construct a stochastic SN equation by adding the stochastic part from
(\ref{stod2}) as a perturbation to the SN equation. Thus our new stochastic SN equation is
\begin{equation}
i\hbar \frac{\partial \Psi({\bf r}, t)}{\partial t} =  - \frac{\hbar^2}{2m} \nabla^{2}\Psi ({\bf r}, t) -
Gm^2 \int \frac{ |\Psi ({\bf r'},t)|^2}{|{\bf r'}-{\bf r}|} d^3 r' \Psi ({\bf r},t) + \left[m\int |\psi_{{\bf r}}({\bf r'},t)|^2\; \phi ({\bf r'},t)\ d^3 r' \right] \Psi ({\bf r}, t)
\label{ssn}
\end{equation}
Here, $\psi_{{\bf r}}$ is a solution of the free wave equation which does not involve the stochastic part, nor the SN potential. The physical interpretation of our analysis will be that the stochastic part decoheres two gaussian wave-packets, whereas the SN part drives the decohered gaussian state to a state with gravitationally inhibited dispersion. In this way is decoherence brought into the SN equation as an additional feature.

The new potential is evidently given by
\begin{equation}
V({\bf r},t) =  -
Gm^2 \int \frac{ |\Psi ({\bf r'},t)|^2}{|{\bf r}-{\bf r'}|} d^3 r'  + m\int |\psi_{{\bf r}}({\bf r'},t)|^2\; \phi ({\bf r'},t)\ d^3 r' 
\label{newpot}
\end{equation}
%It may appear that the second term in the potential does not depend on the position variable ${\bf r}$. However, in principle, by comparison with Di\'osi's mass configuration $f({\bf r}|X)$ one attaches a `configuration variable' $X$ with the probability density $|\Psi({\bf r'},t)|^2$, which should hence be read as
%$|\Psi({\bf r'},t,X)|^2$. Subsequently, when calculating a quantity such as the two point correlation 
%$\langle V({\bf r_1},t), V({\bf r_2},t)\rangle$ for two given points ${\bf r_1}$ and ${\bf r_2}$, one would substitute $X={\bf r_1}$ or $X={\bf r_2}$ respectively. Thus for instance if the quantum state is a gaussian centered at ${\bf r_1}$ (${\bf r_2})$, then $X= {\bf r_1}$ (${\bf r_2})$. This is demonstrated explicitly in the calculation with gaussian states in the next section.

Given the above stochastic potential, we define a stochastic phase $\Phi_{st}({\bf r}, T)$ as follows,
\begin{equation}
\Phi_{st} = - \frac{1}{\hbar} \int_0^T V({\bf r}, t')\; dt'
\label{stopha}
\end{equation}
by integrating the stochastic potential over a time interval $T$. Next we consider the stochastic variance of the difference in this phase at two spatial points ${\bf r_1}$ and ${\bf r_2}$:
\begin{equation}
\Delta \Phi^2 = \langle[\Phi_{st}({\bf {r_1}},t) - \Phi_{st}({\bf r_2},t)]^2\rangle
\label{variance}
\end{equation}
As has been argued by Karolyhazy, and also discussed by us in some detail in a recent work \cite{Bera:2015b}, this phase
variance can be used to test for gravitational decoherence. 
We find the time $T$ for which this variance  is of the order  $\sim \pi^2$ and that will give us the gravitational damping time $T$ for the pair of points ${\bf r_1}$ and ${\bf r_2}$. This method is equivalent to, and 
sometimes simpler than, calculating the damping time from the master equation for the density matrix \cite{Bera:2015}.

We will now calculate this phase variance in order to demonstrate gravitational decoherence, by making an
approximation on the right hand side of the stochastic equation (\ref{ssn}). Namely, we calculate the solution to (\ref{ssn}) iteratively, by replacing the quantum state $\Psi$ on the right hand side of (\ref{ssn}) by the solution $\psi_{\bf r}$ of the free Schr\"{o}dinger equation. By free is meant that we set $V=0$ so that $\psi$ is the solution of the Schr\"{o}dinger  equation for a free particle (no gravitational back-reaction). Thus $V({\bf r},t)$ in (\ref{newpot}) is approximated by replacing $\Psi$ by $\psi_{\bf r}$, and the phase variance is calculated with this approximated potential:
\begin{equation}
V({\bf r},t) =  -
Gm^2 \int \frac{ |\psi _{\bf r}({\bf r'},t)|^2}{|{\bf r}-{\bf r'}|} d^3 r'  + m\int |\psi_{\bf r}({\bf r'},t)|^2\phi({\bf r'},t) \; d^3 r'
\label{newpot2}
\end{equation}

In the next section we demonstrate gravitational decoherence by asking if the particle can be simultaneously in two different gaussian states, one peaked at ${\bf r_1}$, and another peaked at ${\bf r_2}$, and by calculating the phase variance (\ref{variance}) for this pair of states. The respective gaussian peaks  ${\bf r_1}$ and ${\bf r_2}$ serve as `equivalents' of the center of mass of an extended object, and the width of the gaussian serves as equivalent of size of the extended object.

\section{Phase variance and a decoherence criterion for Gaussian states}
Let us assume that the particle has been prepared in an initial state which is a superposition of two gaussian wave-packets, both having the initial width $a$, and one peaked at ${\bf r} = {\bf r_1}$ and the other peaked at  ${\bf r} = {\bf r_2}$. These being solutions of the free particle Schr\"{o}dinger equation, they evolve with time as
\begin{equation}
\psi_{\bf r_1}({\bf r'},t) = (\pi a^2)^{-3/4} \Big(1+\frac{i\hbar t}{ma^2}\Big)^{-3/2}\exp\Big(-\frac{|{\bf r'}-{\bf r_1}|^2}{2a^2(1+\frac{i\hbar t}{ma^2})}\Big)
\label{g1}
\end{equation}
and
\begin{equation}
\psi_{\bf r_2}({\bf r''},t) = (\pi a^2)^{-3/4} \Big(1+\frac{i\hbar t}{ma^2}\Big)^{-3/2}\exp\Big(-\frac{|{\bf r''}-{\bf r_2}|^2}{2a^2(1+\frac{i\hbar t}{ma^2})}\Big)
\label{g2}
\end{equation}
They have the corresponding probability densities
\begin{equation}
|\psi_{\bf r_1}({\bf r'},t)|^2 = \frac{1}{(\pi C_1)^{3/2}} \exp \left ( - \frac {|{\bf r'} - {\bf r_1}|^2}{C_1}\right)
\label{prob1}
\end{equation}
and
\begin{equation}
|\psi_{\bf r_2}({\bf r''},t)|^2 = \frac{1}{(\pi C_1)^{3/2}} \exp \left ( - \frac {|{\bf r''} - {\bf r_2}|^2}{C_1}\right)
\label{prob2}
\end{equation}
where
\begin{equation}
C_1 = a^2 \left(1 + \frac{\hbar ^2 t^2}{m^2 a^4}\right)
\label{C1}
\end{equation}
For each gaussian one calculates the stochastic potential given by Eqn. (\ref{newpot2}) and uses it to calculate the stochastic phase given by Eqn. (\ref{stopha}) thus obtaining expressions for 
$\Phi_{st}({\bf r_1},t)$ and $\Phi_{st}({\bf r_2},t)$. These are then used to obtain the following expression for the phase variance defined in Eqn. (\ref{variance}):
\begin{equation}
\Delta\Phi^2 = I_1 + I_2 + I_3 + I_4 + I_5 + I_6
\end{equation}
where the six integrals $I_1$ to $I_6$ are given by
\begin{equation} 
I_1= \frac{G^2 m^4}{\hbar^2} \int{\frac{ |\psi_{{\bf r_1}}({\bf r'},t')|^2} {{\bf |r_1}-{\bf r'}|}\; \frac{ |\psi_{{\bf r_1}}({\bf r''},t'')|^2} {|{\bf r_1}-{\bf r''}|}} \; d^3r' \; d^3r''\; dt' \;dt''
\label{I1}
\end{equation}
\begin{equation} 
I_2=\frac{G^2 m^4}{\hbar^2} \int{\frac{ |\psi_{{\bf r_2}}({\bf r'},t')|^2} {{\bf |r_2}-{\bf r'}|}\; \frac{ |\psi_{{\bf r_2}}({\bf r''},t'')|^2} {|{\bf r_2}-{\bf r''}|}} \; d^3r' \; d^3r''\; dt' \;dt''
\label{I2}
\end{equation}
\begin{equation} 
I_3=-\frac{2G^2 m^4}{\hbar^2} \int{\frac{ |\psi_{{\bf r_1}}({\bf r'},t')|^2} {{\bf |r_1}-{\bf r'}|}\; \frac{ |\psi_{{\bf r_2}}({\bf r''},t'')|^2} {|{\bf r_2}-{\bf r''}|}} \; d^3r' \; d^3r''\; dt' \;dt''
\label{I3}
\end{equation}
\begin{equation} 
I_4= \frac{ m^2}{\hbar^2} \int \langle\phi({\bf r'},t') \phi({\bf r''},t'')\rangle\; | \psi_{{\bf r_1}}({\bf r'},t')|^2\; 
|\psi_{{\bf r_1}}({\bf r''},t'')|^2 \; d^3r' \; d^3r'' \; dt' \; dt''  
\label{I4}
\end{equation}
\begin{equation} 
I_5= \frac{ m^2}{\hbar^2} \int \langle\phi({\bf r'},t') \phi({\bf r''},t'')\rangle\; | \psi_{{\bf r_2}}({\bf r'},t')|^2\; 
|\psi_{{\bf r_2}}({\bf r''},t'')|^2 \; d^3r' \; d^3r'' \; dt' \; dt''  
\label{I5}
\end{equation}
\begin{equation} 
I_6= - \frac{2 m^2}{\hbar^2} \int \langle\phi({\bf r'},t') \phi({\bf r''},t'')\rangle\; | \psi_{{\bf r_1}}({\bf r'},t')|^2\; 
|\psi_{{\bf r_2}}({\bf r''},t'')|^2 \; d^3r' \; d^3r'' \; dt' \; dt''  
\label{I6}
\end{equation}
This expression for the variance has been arrived at while noting that the stochastic potential has zero mean: $\langle \phi({\bf r},t)\rangle=0$. The first three integrals arise from the non-stochastic SN part of the potential in Eqn. (\ref{newpot2}),  whereas the last three integrals depend on the two point correlation of the stochastic potential, as given in Eqn. (\ref{dcorr}).

These integrals are then evaluated for the gaussian states (\ref{g1}) and \ref{g2}). Details are given in Appendix I. It is easily shown that the results of the first three integrals are independent of ${\bf r_1}$ and ${\bf r_2}$ and that these three integrals cancel each other exactly. This is understandable because the SN part, being deterministic,  should not contribute to the phase variance and to decoherence. After carrying out the spatial integration in the last three integrals one gets the following result for the phase variance
\begin{equation}
\Delta\Phi^2 = \frac{2\sqrt{2}\kappa}{\sqrt \pi} \int_0^T \frac{1}{\sqrt {C_1}} \; dt
-\frac{2\kappa}{R} \int _0^T {\rm Erf}\left(\frac{R}{\sqrt{2C_1}}\right)\; dt  \quad \quad \ \equiv I_7 - I_8
\label{phavar}
\end{equation}
where $\kappa = Gm^2/\hbar$, $R=|{\bf r_1}-{\bf r_2}|$ and Erf is the error function, and $C_1$ is as defined in (\ref{C1}).

The first of these integrals, $I_7$, is easily carried out and the result is
\begin{equation}
I_7 = \frac{2\sqrt{2}\kappa}{\sqrt{\pi}} \frac{ma}{\hbar} \sinh^{-1}\left(\frac{\hbar T}{ma^2}\right) =
 \frac{2\sqrt{2}\kappa}{\sqrt{\pi}} \frac{ma}{\hbar} \; \left[ \frac{\hbar T}{ma^2} - 
 \frac{1}{6}\left( \frac{\hbar T}{ma^2}\right)^3 + {\cal O}\left( \frac{\hbar T}{ma^2}\right)^5       \right]
 \label{I7}
\end{equation} 
The second integral $I_8$ involving the error function cannot be carried out exactly, and to begin with we carry out the integral in the limit, $T\ll ma^2/\hbar$. In this limit we have $C_1 = a^2 \left(1 + \frac{\hbar ^2 t^2}{m^2 a^4}\right) \approx a^2$ and the argument of the Erf function in $I_8$ is approximately a constant $R/\sqrt{2}a$. The integral then becomes trivial, and after retaining only the leading order term in the expansion of $I_7$ in (\ref{I7}), and setting the phase variance to be of the order $\pi^2$ in Eqn. (\ref{phavar}) we arrive at the following expression for the gravitational damping time $T$ for two gaussian wave-packets which are separated by a distance $R$:
\begin{equation}
T^{-1} \sim \frac{Gm^2}{\hbar} \left[ \sqrt{\frac{2}{\pi}} \frac{1}{a} -  \frac{1}{R} \rm{Erf} \left(\frac{R}{\sqrt{2}a}\right)   \right] 
\end{equation}

With the superposed state formed from the two gaussian wave-packets we associate a coherence length 
$L$ and the quantum kinematic time scale $t_q \sim mL^2/\hbar$. $L$ is that characteristic separation distance $R$ beyond which the off-diagonal elements of the density matrix become negligibly small.   If $t_q > T$ then gravitational damping is effective and the two states decohere. If $T > t_q$ then decoherence is ineffective. We can define the critical length $L_c$ as the length for which these two time scales become equal: $T=t_q$. For $L > L_c$ decoherence is effective, and for $L < L_c$ it is not effective. The critical length $L_c$ is hence given by the relation
\begin{equation}
t_q^{-1} \sim\frac{\hbar}{mL_c^2}\sim T^{-1}(L_c) \sim \frac{Gm^2}{\hbar} \left[ \sqrt{\frac{2}{\pi}} \frac{1}{a} -  \frac{1}{L_c} \rm{Erf} \left(\frac{L_c}{\sqrt{2}a}\right)   \right] 
\end{equation}
Since this expression for damping time is valid only when $T\ll ma^2/\hbar$ it is evident that we are considering the case $L_c \ll a$. Under this condition the term with Erf function can be approximated as
\begin{equation}
\frac{1}{L_c} {\rm erf} \left( \frac{L_c}{\sqrt{2} a}\right) =  \sqrt{\frac{2}{\pi}} \frac{1}{a}
- \frac{2L_c^2}{6\sqrt{2\pi}a^3}
\end{equation}
so that the critical length is given by
\begin{equation}
L_c \sim \left( \frac{\hbar^2}{Gm^3}\right)^{1/4} a^{3/4},     \qquad   {\rm if} \qquad am^3 \gg \hbar^2 / G
\label{macro}
\end{equation}
This is the same result as obtained by Di\'osi \cite{Diosi:87} using a classical mass distribution, whereas we have used a quantum probability density for a gaussian wave-packet, instead.

Let us now consider the opposite extreme $T\gg ma^2/\hbar$, $R\gg a$ for evaluating the integrals in
(\ref{phavar}). The Erf integral still cannot be evaluated exactly, and we will obtain an approximate expression for the critical length, by examining the behaviour of the integrands. To do so, we define the quantity 
\begin{equation}
\beta(t) =  \frac{R}{\sqrt{2} a} \frac{1} {\sqrt{ 1+ \hbar^2 t^2 /m^2 a^4 } }
\end{equation}
in terms of which the expression (\ref{phavar}) for the phase variance can be rewritten as follows:
\begin{equation}
\Delta\Phi^2 = \frac{4\kappa }{\sqrt \pi R}  \int_0^T \; dt\; \left [ \beta (t) - 
\int_0^{\beta (t)} \; dx\; \exp (-x^2) \right ]\quad  \equiv \quad \frac{4\kappa }{\sqrt \pi R}  \int_0^T \; dt\; I(t)
\label{integ}
\end{equation}

Writing it thus facilitates a comparison between the first term $\beta (t)$ and the second term which is an Erf integral whose upper limit is $\beta (t)$.  At the lower limit $t=0$ of the time integration,  we have that $\beta(0) = R/\sqrt{2} a \gg 1$. Now the error function is bounded from above by unity, so at $t=0$ the Erf term can be ignored compared to the first term $\beta (t)$, and the integrand in (\ref{integ}) takes the initial value   $I(t=0)\sim R/a \gg 1$. At the upper limit $t=T$ we have $T\gg ma^2 /\hbar$ and hence 
$\beta(T) \sim (R/a) (ma^2 / \hbar T)$ which is a product of two quantities, the first of which satisfies $R/a \gg 1$ and the second of which satisfies $ma^2 / \hbar T \ll 1$. It can be argued that their product is much less than unity. To see this, we write $T$ in terms of the critical length $L_c$ as $T\sim mL_c^2/\hbar$, hence getting $\beta(T) \sim Ra / L_c^2$. Since eventually $R$ is set equal to $L_c$, we get that $\beta(T)= a/L_c \ll 1$ and hence $\beta(T)\sim 0$. 

Since we have obtained $\beta(0) \sim R/a \gg 1$ and $\beta(T) \sim 0$, a very rough approximation to the integral (\ref{integ}) is to say that the integral is of the order of the range $T$ of the integral, multiplied by the difference $R/a$ between the values of the integrand $I(t)$ at the upper end and the lower end. Thus we get that the integral is of the order $T. R/a$ and this suffices for our purpose. We hence get that
\begin{equation}
T \sim \frac{\hbar a}{Gm^2}, \qquad L_c \sim\left(  \frac{\hbar ^2}{Gm^3} \right)^{1/2} a^{1/2},   \qquad   {\rm if} \qquad am^3 \ll \hbar^2 / G
\label{micro}
\end{equation}
This is the same result as obtained by Di\'osi \cite{Diosi:87} in the microscopic limit.

 \section{Discussion}
 In this paper we have proposed that the Di\'osi correlation (\ref{dcorr}) should be used to make a stochastic modification of the SN equation, by replacing the classical mass distribution used in (\ref{stod}) by the quantum probability density. This allows for a joint consideration of self-gravity effects as well as the effect of extrinsic spacetime fluctuations. Interestingly, by using gaussian wave-packets we find the same results for gravitational decoherence as found in \cite{Diosi:87} using a classical mass distribution. The analog of the size of a classical object is here the initial width $a$ of the gaussian, whereas the separation $R$ between the peaks of the two gaussians is the analog  of separation between two classical positions of an object. It seems to us that since one cannot meaningfully set $a$ to zero, one naturally avoids the point-particle divergence which arises while doing the analysis using classical mass distributions.
 
 Considering that the SN part does not contribute to gravitational decoherence, one could well ask if we have genuinely found a new stochastic SN equation, or have we merely recovered Di\'osi's results by employing wave-packets? We are inclined to believe the former, because the use of a quantum probability density in the stochastic modification does not seem to make sense except in the SN context. Because, if the quantum probability density couples to the stochastic potential, there is no reason why it should not act as a source for self-gravity as well.  Thus in our view, (\ref{ssn}) is a sensible equation for studying gravitational decoherence. In particular, the new equation overcomes the problem of `wrong Newtonian limit' \cite{Bahrami:2014} faced by the SN equation, because Newtonian limit should now be examined at the level of the stochastic equation, where we do find decoherence and selection of one or the other wave-packet, which represents the correct Newtonian limit. By wrong Newtonian limit we mean that a superposition of two Gaussians would "collapse" to the center of the two rather than each of the two positions with 50\% probability. This of course violates the Born rule as well. This problem is avoided in our current stochastic treatment. 
 
 The new stochastic equation does have a limitation. It continues to possess the nonlinear structure of the SN equation, as a result of which the master equation  for the density matrix has a complicated nonlinear structure which does not prevent the superluminality problem of the SN equation. However, in the approximation under which the present analysis has been done, namely the use of the free solution in the approximated potential (\ref{newpot2}), the master equation is linear and of the Lindblad form, exactly the same as in \cite{Diosi:87},  with the mass density replaced by the quantum probability density constructed from the free state. At this level of approximation, superluminal signalling is avoided in the stochastic SN equation. 
 
 Our results match with the numerical estimates for the quantum--classical transition, as given in \cite{Diosi:87}, once we identify the gaussian width $a$ with the `size' of the object. In particular, we know that the critical quantum-classical transition is obtained by setting the critical length $L_c$ such that $L_c\sim a$.
 If we put $L_c=a$ in (\ref{macro}) or (\ref{micro}) we get $L_c=a\sim \hbar^2 / Gm^3$, a well-known result.
 [This length scale also emerges when one considers gravitationally induced dispersion of a wave-packet in the SN equation.
  \cite{giulini2011gravitationally}].
Assuming $a$ to be the size of the object, and using $m= 4\pi  a^3 \rho/3$ where $\rho$ is the density,
we get that
\begin{equation}
m_c \sim \left ( \frac {\hbar \rho^{1/6}} {G^{1/2}}\right)^{3/5}
\end{equation}
For a density of 1 gm/cc we get $m_c \sim 10^{-14}$ g. Therefore, we conclude that if $m>m_c$ then $L_c < a$ and the object is classical. If $m<m_c$ then $L_c > a$ and the object is quantum.

Stochastic SN equations have been proposed earlier as well, at least on two occasions, once in  \cite{nimmrichter:2015} and in another paper, by us \cite{Bera:2015b}. We proposed to find stochastic modifications to the SN equation by studying higher order stochastic corrections to semiclassical gravity and taking the Newtonian limit of the theory. The difference between that treatment and the present one is that there the stochastic corrections are due to self-gravity, whereas in the present work these are coming due to extrinsic spacetime fluctuations. It would be hard to say which treatment is preferred over the other, although the present work has the merit of exactly reproducing earlier decoherence results. In principle, it would appear that both self-gravity stochastic corrections    as well as extrinsic fluctuations should be considered. Our comments on the interesting work of \cite{nimmrichter:2015} can be found in \cite{Bera:2015b}. 

The study of stochastic SN equations has a long way to go, before the equation can [if at all] acquire the appealing structure of the CSL model. However, since no theoretical underpinning for the phenomenological CSL model is known, and since gravity is a universal interaction, it is worthwhile to explore if collapse models can be realised by generalising models of gravitationally induced decoherence.

\bigskip

\noindent{\bf Acknowledgements:} We would like to thank Lajos Di\'osi for his critical comments which significantly improved an earlier version of the manuscript. PG would like to thank the Tata Institute of Fundamental Research, where this work was carried out, for its kind hospitality during her visit.

\bigskip
 
 \section{Appendix I} 
 Evaluation of the integrals $I_1$ to $I_6$ given by Eqns. (\ref{I1}) to (\ref{I6}):
 
{\bf Integral $I_1$ :} By substituting the probability density (\ref{prob1}) in the expression (\ref{I1}) for the integral $I_1$ we can write this integral as
 \begin{equation}
 I_1 = \frac{\kappa^2}{\pi} \int{\frac{1}{C_1^3}\frac{\exp \left ( - \frac {|{\bf r'} - {\bf r_1}|^2}{C_1}\right)}{{\bf |r_1}-{\bf r'}|}}\; \frac{\exp \left ( - \frac {|{\bf r''} - {\bf r_1}|^2}{C_1}\right)} {|{\bf r_1}-{\bf r''}|} \; d^3r' \; d^3r''\; dt' \;dt''
 \end{equation}
 The substitutions ${\bf R'}={\bf r_1} - {\bf r'}$  and ${\bf R''}={\bf r_1} - {\bf r''}$ transform the integral to
 \begin{equation}
 I_1 = \frac{\kappa^2}{\pi} \int \frac{1}{C_1^3}\frac{\exp \left ( - \frac {R'^2 + R''^2}{C_1}\right)}{R'R''}\; d^3R' \; d^3R''\; dt' \;dt''
 \end{equation}
 
 {\bf Integral $I_2$ :} The substitutions   ${\bf R'}={\bf r_2} - {\bf r'}$  and ${\bf R''}={\bf r_2} - {\bf r''}$ bring the integral $I_2$ given by (\ref{I2}) to the same form as $I_1$ above, so that $I_1$=$I_2$.
 
 {\bf Integral $I_3$ :} Similarly, the substitutions   ${\bf R'}={\bf r_1} - {\bf r'}$  and ${\bf R''}={\bf r_2} - {\bf r''}$ in (\ref{I3}) show that $I_3 = -2I_1 = -2I_2$ 
 
 Hence it follows that $I_1+I_2+I_3=0$.  As noted earlier, this result is expected because the SN part should not contribute to the variance.
 
{\bf  Integral $I_4$ :} Using the form of the correlation function given by (\ref{dcorr}) and the gaussian probability density given by (\ref{prob1}) the spatial part of the  integral (\ref{I4}) can be written as 
 \begin{equation}
 I_{4s} = \frac{\kappa}{(\pi C_1)^3} \int \frac{ \exp\left  (-\frac{z'^2}{C_1}\right) \exp\left  (-\frac{z''^2}{C_1}\right)} { |{\bf z'}-{\bf z''}|} \; d^3 z' \; d^3 z''
 \end{equation}
 where ${\bf z'} = {\bf r'} - {\bf r_1}$ and ${\bf z''} = {\bf r''} - {\bf r_1}$. 
 By writing ${\bf z'}$ in spherical polar coordinates $(z',\theta,\phi)$ and by taking $\theta=0$ along ${\bf z''}$
 this integral can be written as
  \begin{equation}
 I_{4s} = \frac{\kappa}{(\pi C_1)^3} \int \frac{ \exp\left  (-\frac{z'^2}{C_1}\right) \exp\left  (-\frac{z''^2}{C_1}\right)} { \sqrt{z'^2 + z''^2 - 2z'z''\cos\theta}} \;  z'^2 dz' \sin\theta\; d\theta\; d\phi \; d^3 z''
 \end{equation}
 and after carrying out the $\theta$ integral, followed by the trivial angular integrals in ${\bf z''}$ we get
  \begin{equation}
 I_{4s} = \frac{\kappa}{(\pi C_1)^3} (2\pi)^2.2\int  \exp\left  (-\frac{z'^2}{C_1}\right) \exp\left  (-\frac{z''^2}{C_1}\right) \left [ z'+z'' - |z'-z''|\right ]\; z'\; z'' \; dz' \; dz'' \equiv I_{4sa}+I_{4sb}
 \end{equation}
 Here, $I_{4sa}$ is the integral for $z'<z''$ and is given by
 \begin{equation}
 I_{4sa} = \frac{\kappa}{(\pi C_1)^3} (2\pi)^2.2\int _{z''=0}^{\infty} \exp\left  (-\frac{z''^2}{C_1}\right) \; z'' dz''
 \int_{0}^{z''}  \exp\left  (-\frac{z'^2}{C_1}\right) \; 2z'^2 \; dz'
 \end{equation}
  while $I_{4sb}$ is the integral for $z>z''$ and is given by
 \begin{equation}
 I_{4sb} = \frac{\kappa}{(\pi C_1)^3} (2\pi)^2.4\int _{z''=0}^{\infty} \exp\left  (-\frac{z''^2}{C_1}\right) \; z''^2 dz''
 \int_{z=z''}^{\infty}  \exp\left  (-\frac{z'^2}{C_1}\right) \; z' \; dz'
 \end{equation}

The integrals $I_{4sa}$ and $I_{4sb}$ are easily carried out, giving the final simple result that
\begin{equation}
I_{4s} = \sqrt{\frac{2}{\pi}} \; \kappa \; \frac{1} {\sqrt{C_1}}
\end{equation}
and hence, after including the time integral, we get
\begin{equation}
I_4 =  \frac{\sqrt{2}\kappa}{\sqrt \pi} \int_0^T \frac{1}{\sqrt {C_1}} \; dt
\end{equation}

{\bf Integral $I_5$ :} This integral, given by (\ref{I5}), is easily seen to be equal to $I_4$, and hence the sum $I_4+I_5$ gives the first integral on the right side of the expression (\ref{phavar}) for the variance. The second integral in (\ref{phavar}) comes from the $I_6$ of (\ref{I6}).

{\bf Integral $I_6$ :} Using the form of the correlation function given in (\ref{dcorr}), along with the expressions (\ref{prob1}) and (\ref{prob2}) for the probability density, and the definitions  ${\bf z'} = {\bf r'} - {\bf r_1}$ and ${\bf z''} = {\bf r''} - {\bf r_1}$ introduced above, the spatial part of the integral (\ref{I6}) can be written as
\begin{equation}
 I_{6s} = -\frac{2\kappa}{(\pi C_1)^3} \int \frac{ \exp\left  (-\frac{z'^2}{C_1}\right) \exp\left  (-\frac{|{\bf z''}+{\bf R}|^2}{C_1}\right)} { |{\bf z'}-{\bf z''}|} \; d^3 z' \; d^3 z''
 \end{equation}
where ${\bf R} = {\bf r_1} - {\bf r_2}$. As before, we introduce spherical polar coordinates $(z',\theta,\phi)$ for ${\bf z'}$ and by taking $\theta=0$ along ${\bf z''}$ and carrying out the $\theta$ integral we get
 \begin{equation}
 I_{6s} =- \frac{2\kappa}{(\pi C_1)^3} (2\pi)\int  \frac{1}{z'z''}\exp\left  (-\frac{z'^2}{C_1}\right) \exp\left  (-\frac{|{\bf z''}+{\bf R}|^2}{C_1}\right) \left [ z'+z'' - |z'-z''|\right ]\; z'^2\;  dz' \; d^3z'' 
 \end{equation}
 To address the presence of the vector ${\bf R}$, introduce spherical polar coordinates ${(z'',\theta'',\phi'')}$ for ${\bf z''}$ with the $\theta''=0$ axis aligned along ${\bf R}$, so that $\theta''$ is the angle between ${\bf z''}$ and ${\bf R}$. 
 Then by writing $d^3z''=z''^2 \sin\theta''d\theta'' d\phi'' dz''$ and by first carrying out the angle integral
 \begin{equation}
 X\equiv \int_{0}^{\pi}  \exp\left  (-\frac{|{\bf z''}+{\bf R}|^2}{C_1}\right) \sin\theta'' \; d\theta''=
  \int_{0}^{\pi}  \exp\left  (-\frac{(z''^2 + R^2 + 2z''R\cos\theta'')}{C_1}\right) \sin\theta'' \; d\theta''
 \end{equation}
 the integral $I_{6s}$ can be written as
 \begin{equation}
 \begin{split}
 I_{6s} =  - \frac{2\kappa}{(\pi C_1)^3} \frac{(2\pi)^2 C_1}{2R} \int  & \exp[ -z'^2/C_1]
 \left[\exp-(z''-R)^2/C_1 - \exp-(z''+R)^2/C_1 \right] \times \\
 & (z'+z''-|z'-z''|) \; z'dz'dz''
  \end{split}
 \end{equation}

This may further be written as the sum $I_{6s}\equiv I_{6sa} + I_{6sb}$ where $I_{6sa}$ is the integral for $z<z'$ and $I_{6sb}$ is the integral for $z>z'$, and these are given by
 \begin{equation}
 %\begin{split}
 I_{6sa} = -  \frac{2\kappa}{(\pi C_1)^3} \frac{(2\pi)^2 C_1}{2R} \int_{z''=0}^{\infty}   
 \left[\exp-(z''-R)^2/C_1 - \exp-(z''+R)^2/C_1 \right] 
 \int_{z'=0}^{z''}   \exp[ -z'^2/C_1] 2z'^2 \; dz'dz''
  %\end{split}
 \end{equation}
and
\begin{equation}
 %\begin{split}
 I_{6sb} =  -\frac{2\kappa}{(\pi C_1)^3} \frac{(2\pi)^2 C_1}{2R} \int_{z''=0}^{\infty}   
 \left[\exp-(z''-R)^2/C_1 - \exp-(z''+R)^2/C_1 \right] 
 \int_{z'=z''}^{\infty}   \exp[ -z'^2/C_1] 2z'z'' \; dz'dz''
  %\end{split}
 \end{equation}
 These last two integrals are easily carried out and together give the result that
 \begin{equation}
 I_{6s} = -\frac{2\kappa}{\sqrt{\pi C_1} R}\int_{0}^{\infty}  \left[\exp-(z''-R)^2/C_1 - \exp-(z''+R)^2/C_1 \right] 
 {\rm Erf}\left(\frac{z''}{\sqrt{C_1}}\right)\; dz''
 \end{equation}
Making the substitution $z''/\sqrt{C_1}=x$ and using the result that [see Eqn. 4.3.37 of \cite{erfinteg}]
\begin{equation}
\int_{0}^{\infty} {\rm Erf}(x) \left\{ \exp\left[-\left(x-\frac{R}{\sqrt{C}}\right)^2\right] - \exp\left[-\left(x+\frac{R}{\sqrt{C}}\right)^2\right] \right\}\; dx = \sqrt{\pi} {\rm Erf} \left(\frac{R}{\sqrt{2C_1}}\right)
\end{equation}
we get that
\begin{equation}
I_{6s} = -\frac{2\kappa}{R}  {\rm Erf} \left(\frac{R}{\sqrt{2C_1}}\right)
\end{equation}  
The time integral of this spatial part appears as the second integral on the right hand side of (\ref{phavar}). 

\bigskip
\centerline{\bf REFERENCES}

\bibliography{biblioqmtstorsion}

\end{document}